\documentclass[11pt]{article}
\usepackage[utf8]{inputenc}
\usepackage{jheppub}
\usepackage{graphicx}
\usepackage{caption}
\usepackage{subcaption}
\usepackage{hyperref}
\usepackage[T1]{fontenc}
\usepackage{float}

\newcommand{\be}{\begin{equation}}
\newcommand{\ee}{\end{equation}}
\newcommand{\bea}{\begin{eqnarray}}
\newcommand{\eea}{\end{eqnarray}}

\newcommand\underrel[3][]{\mathrel{\mathop{#3}\limits_{%
      \ifx c#1\relax\mathclap{#2}\else#2\fi}}}
\title{The five-point bootstrap}

\author{David Poland,$^{a}$ Valentina Prilepina,$^{b}$ Petar Tadi\' c$^{a}$}

\affiliation{$^{a}$ Department of Physics, Yale University, New Haven, CT 06520, USA}
\affiliation{$^{b}$ Perimeter Institute for Theoretical Physics, Waterloo, ON N2L 2Y5, Canada}

\emailAdd{david.poland@yale.edu, valentina.prilepina.1@ulaval.ca, petar.tadic@yale.edu}

\abstract{We study five-point correlation functions of scalar operators in  $d$-dimensional conformal field theories. We develop a new approach to computing the five-point conformal blocks for exchanged primary operators of arbitrary spin by introducing a generalization of radial coordinates, using an appropriate ansatz, and perturbatively solving two quadratic Casimir differential equations. We then study five-point correlators $\langle\sigma\sigma\epsilon\sigma\sigma\rangle$ in the critical 3d Ising model. We truncate the operator product expansions (OPEs) in the correlator by including a finite number of primary operators with conformal dimension below a cutoff $\Delta\leqslant \Delta_{\rm cutoff}$. We then compute several OPE coefficients involving $\epsilon$ and two spinning operators by demanding that the truncated correlator approximately satisfies the crossing relation.}

\begin{document}
\maketitle
\flushbottom

\newpage
\section{Introduction}

Conformal field theories (CFTs) are ubiquitous in modern physics. They play a very special role in the space of all quantum field theories as they represent fixed points of renormalization group flows. They also describe second-order phase transitions in condensed matter systems and furnish a  useful handle on black hole physics and quantum gravity through the AdS/CFT correspondence. In recent years, much remarkable progress has been made in understanding the space of CFTs and their dynamics by use of the conformal bootstrap, following \cite{Ferrara:1973yt, Polyakov:1974gs, Rattazzi:2008pe, El-Showk:2012cjh}. The conformal bootstrap is a non-perturbative method for solving CFTs by applying the implications of conformal symmetry and imposing consistency conditions. In this context, to solve a  conformal field theory means to determine the spectrum of scaling dimensions of primary operators and their three-point functions, which are fixed by conformal symmetry up to a finite number of constants, also known as the operator product expansion (OPE) coefficients.

So far, the conformal bootstrap has mostly been applied to four-point correlation functions with external scalar operators (see~\cite{Iliesiu:2015qra, Iliesiu:2017nrv, Dymarsky:2017xzb, Dymarsky:2017yzx, Karateev:2019pvw, Reehorst:2019pzi, Erramilli:2020rlr, Erramilli:2022kgp, He:2023ewx} for some exceptions). Therefore, most of the known OPE coefficients involve two scalars and one spinning operator. OPE coefficients involving multiple spinning operators are typically not known. One way to obtain these OPE coefficients is to consider four-point correlation functions with external spinning operators, but this approach is technically challenging. Another way is to impose consistency conditions on higher-point correlation functions with external scalar operators. Here the difficulty is that the higher-point conformal blocks are generally not available in an explicit form.

One notable exception is the conformal block for exchanged scalar operators in five-point correlation functions, which was computed in \cite{Rosenhaus:2018zqn}. The higher-point conformal blocks for exchanged scalar operators were also computed using holographic methods in \cite{Parikh:2019ygo, Parikh:2019dvm, Hoback:2020pgj, Fortin:2022grf} and by means of dimensional reduction in \cite{Hoback:2020syd}. 
A relation between higher-point conformal blocks and solutions of a Lauricella system was established in \cite{Pal:2020dqf}, while a connection to Gaudin models was made in \cite{Buric:2020dyz, Buric:2021ywo, Buric:2021ttm, Buric:2021kgy}.
General representations of higher-point conformal blocks were obtained using the operator product expansion in embedding space in \cite{Skiba:2019cmf, Fortin:2019dnq, Fortin:2019zkm, Fortin:2020ncr, Fortin:2020yjz, Fortin:2020bfq}. Beyond scalar exchange, five-point conformal blocks for exchanged spinning operators (with identical external scalars) were computed in \cite{Goncalves:2019znr}. However, explicit evaluation of these blocks is technically cumbersome as the expressions involve summing over 9 variables, with coefficients that must be determined recursively by solving the Casimir differential equations. Explorations of the conformal bootstrap for higher-point correlation functions have been initiated in \cite{Bercini:2020msp, Antunes:2021kmm, Anous:2021caj, Kaviraj:2022wbw, Goncalves:2023oyx}.

Recently, recursion relations that relate the five-point conformal blocks for exchanged operators of different spin were explicitly derived in \cite{Poland:2021xjs}. By solving these recursion relations one can express the five-point conformal block for the exchanged operators of arbitrary spin as a linear combination of five-point conformal blocks for exchanged scalar operators with shifted conformal dimensions. The five-point conformal blocks for the exchanged scalar operators can be evaluated using \cite{Rosenhaus:2018zqn, Parikh:2019dvm, Fortin:2019zkm}. Therefore, following this approach, it is straightforward to compute five-point conformal block for any spin of exchanged operators. While this approach works, it is fairly cumbersome at higher spins, due to the large number of terms generated by the recursion relation.

In this paper we first present an alternate method for computing the five-point conformal blocks with arbitrary spins exchanged by introducing a generalization of the radial coordinates developed for four-point conformal blocks in~\cite{Hogervorst:2013sma}. We then write down a general series expansion for the conformal blocks in these coordinates and compute the coefficients in this expansion by solving the two quadratic Casimir differential equations satisfied by these objects. This approach is similar in spirit to the strategy employed in~\cite{Goncalves:2019znr}, but the resulting structure is greatly simplified due to our choice of coordinates. As a demonstration of its utility, we compute the OPE coefficients in mean-field theory (MFT) involving two spinning operators by expanding the MFT five-point function in terms of our conformal blocks found using this technique.

We then use these conformal blocks to numerically impose consistency conditions on five-point correlation functions in the 3d free scalar theory and the critical 3d Ising model. In this work, our goal is to compute the OPE coefficients of the latter involving one scalar and two spinning operators. These OPE coefficients are for example inputs for the Hamiltonian truncation technique, which is another non-perturbative method for studying the dynamics of the strongly coupled quantum field theories, see e.g.~\cite{Fitzpatrick:2022dwq}. Unlike in the usual four-point case, there is no positivity in the expansion of the five-point correlation function (so that convex optimization methods cannot be easily used); instead, these coefficients can be approximately computed by truncating the OPEs and minimizing a cost function which measures how close the crossing relations are to being satisfied.

The OPE coefficients of one scalar and two spinning operators in the critical Ising model that we compute with reasonable accuracy are given by  
\begin{equation}
\begin{split}
\lambda^{0}_{T\epsilon T} &\approx 0.81(5),\\
\lambda^{0}_{T\epsilon C} &\approx 0.30(6) ,\\    
\lambda^{4}_{C\epsilon C} &\approx -0.3(1),
\end{split}
\end{equation}
where $\epsilon$ is the $\mathbf{Z}_{2}$-even scalar with the lowest conformal dimension, $T$ is the spin-2 stress tensor, and $C$ is the spin-4, $\mathbf{Z}_{2}$-even operator with the lowest conformal dimension. Superscripts in the OPE coefficients denote tensor structures of the corresponding three-point function. We use the standard box basis of the conformally invariant tensor structures in the three-point functions, defined in \cite{Costa:2011mg}. These results are not yet very precise and come from a fairly severe truncation (just up to spin 4), but our takeaway message is that the five-point bootstrap works and can likely be improved by including more operators into the system.

This paper is organized as follows. In section~\ref{sec:5pt}, we demonstrate the new method for computing the five-point conformal blocks for arbitrary spin of the exchanged operators by constructing an appropriate ansatz in the radial coordinates and solving two quadratic Casimir differential equations. We then expand the mean-field theory five-point correlator $\langle\phi\phi[\phi,\phi]_{0,0}\phi\phi\rangle$ in terms of five-point conformal blocks and compute OPE coefficients for all exchanged operators in this five-point function. In section~\ref{sec:truncation}, we truncate the OPEs in the five-point correlators of three-dimensional free theory and the critical Ising model by including just a finite number of exchanged operators of conformal dimension below a certain cutoff $\Delta\leqslant \Delta_{\rm cutoff}$. We then demand that the truncated correlators approximately satisfy the crossing relation in order to numerically compute OPE coefficients of one scalar and two spinning operators. We discuss our results in section~\ref{sec:discussion}.

\section{Five-point conformal blocks}
\label{sec:5pt}

In this section we review five-point functions in $d>2$ dimensions and present a new method for computing the conformal blocks of exchanged operators with arbitrary spin. We then study the five-point correlator $\langle\phi\phi[\phi,\phi]_{0,0}\phi\phi\rangle$ in mean-field theory in arbitrary dimension $d$. In particular, we expand this correlator in terms of the five-point conformal blocks and compute the OPE coefficients of all contributions.

To begin, let us consider a general five-point correlator with external scalar operators 
\begin{equation}
\langle\phi_1(x_1)\phi_2(x_2)\phi_3(x_3)\phi_4(x_4)\phi_5(x_5)\rangle.
\end{equation} Using the operator product expansion in the (12) and (45) channels, the correlator can be expanded as
\begin{equation}\label{first-channel}
\begin{split}
&\langle\phi_1(x_1)\phi_2(x_2)\phi_3(x_3)\phi_4(x_4)\phi_5(x_5)\rangle =\\
&\sum_{(\mathcal{O}_{\Delta,l},\mathcal{O'}_{\Delta',l'})}\sum_{n_{IJ}=0}^{{\rm min}(l,l')} \lambda_{\phi_1\phi_2 \mathcal{O}_{\Delta,l}}\lambda_{\phi_4\phi_5 \mathcal{O'}_{\Delta',l'}}\lambda_{\mathcal{O}_{\Delta,l} \phi_3 \mathcal{O'}_{\Delta',l'}}^{n_{IJ}}P(x_i)G^{(n_{IJ})}_{(\Delta, l, \Delta', l')}(u_1',v_1',u_2',v_2',w').
\end{split}
\end{equation}
The conformal blocks $G^{(n_{IJ})}_{(\Delta, l, \Delta', l')}$ encode the contributions to the five-point correlator of a pair of primary operators $(\mathcal{O}_{\Delta,l},\mathcal{O'}_{\Delta',l'})$ and their descendants that appear in the $\phi_1 \times \phi_2$ and $\phi_4 \times \phi_5$ operator product expansions, respectively. They are completely fixed by conformal symmetry. The sums over $\mathcal{O}_{\Delta,l}$ and $\mathcal{O'}_{\Delta',l'}$ run over all primary operators in the spectrum.  The superscript $n_{IJ}$ labels the independent conformally-invariant tensor structures of the three-point function $\langle \mathcal{O}_{\Delta,l} \phi_3 \mathcal{O'}_{\Delta',l'} \rangle$, with $n_{IJ} = 0,1,\ldots, {\rm min}(l,l')$.

Here we follow the conventions of \cite{Rosenhaus:2018zqn}. The pre-factor function $P(x_i)$ (often referred to as the external leg factor) and the conformal cross-ratios are given by
\begin{equation}\label{cros-rat-one}
\begin{split}
&P(x_i)=\frac{1}{x_{12}^{\Delta_{1}+\Delta_{2}} x_{34}^{\Delta_{3}} x_{45}^{\Delta_{4}+\Delta_{5}}}\left(\frac{x_{23}}{x_{13}}\right)^{\Delta_{12}}\left(\frac{x_{24}}{x_{23}}\right)^{\Delta_{3}}\left(\frac{x_{35}}{x_{34}}\right)^{\Delta_{45}}, \qquad x_{ij}=x_i-x_j,\\
&u_1'=\frac{x_{12}^{2}x_{34}^{2}}{x_{13}^{2}x_{24}^{2}},\quad v_1'=\frac{x_{14}^{2}x_{23}^{2}}{x_{13}^{2}x_{24}^{2}},\quad u_2'=\frac{x_{23}^{2}x_{45}^{2}}{x_{24}^{2}x_{35}^{2}},\quad v_2'=\frac{x_{25}^{2}x_{34}^{2}}{x_{24}^{2}x_{35}^{2}},\quad w'=\frac{x_{15}^{2}x_{23}^{2}x_{34}^{2}}{x_{24}^{2}x_{13}^{2}x_{35}^{2}},
\end{split}
\end{equation}
where $\Delta_{i}$ are conformal dimensions of $\phi_i$ and $\Delta_{ij}=\Delta_{i}-\Delta_{j}$. 

The conformal blocks for exchanged scalar operators $(l=l'=n_{IJ}=0)$ were computed in \cite{Rosenhaus:2018zqn, Parikh:2019dvm, Fortin:2019zkm}. One can then obtain the conformal blocks for the exchanged spinning operators in terms of the scalar exchange blocks by using recursion relations that were derived in \cite{Poland:2021xjs}. In particular, the recursion relations allow one to express the conformal block for exchanged spinning operators as a linear combination of the conformal blocks for the exchanged scalar operators with shifted conformal dimensions of both external and exchanged operators. 

Here, we describe an alternative way to compute the five-point conformal blocks by solving two quadratic Casimir equations, which is similar to the approach of~\cite{Goncalves:2019znr}. As shown by Dolan and Osborn, the conformal blocks are eigenfunctions of the differential Casimir operators of the conformal group \cite{Dolan:2003hv, Dolan:2011dv}. We use the cross-ratios defined in \citep{Buric:2021ywo, Buric:2021kgy}, given by
\begin{equation}
\begin{split}
& u_1'=z_1 \bar{z}_1, \qquad v_1'=(1-z_1)(1-\bar{z}_1),\\
& u_2'=z_2 \bar{z}_2, \qquad v_2'=(1-z_2)(1-\bar{z}_2),\\
& w'=w(z_1-\bar{z}_1)(z_2-\bar{z}_2)+(1-z_1-z_2)(1-\bar{z}_1-\bar{z}_2).
\end{split}
\end{equation}
As explained in~\cite{Buric:2021kgy} (see figure 4), these cross-ratios have a simple geometric interpretation in the following conformal frame. Using conformal transformations, the points $x_2$, $x_4$, and $x_3$ can be placed on a single line at positions $0$, $1$, and $\infty$, respectively. Next, rotations transverse to this line can be used to put $x_1$ on a plane. Using the remaining rotations transverse to this plane we can then place $x_5$ within a three-dimensional subspace. In this subspace, there are two planes $x_1x_2x_3x_4$ and $x_2x_3x_4x_5$ that intersect along the line where the points $x_2$, $x_4$, and $x_3$ are placed. The cross-ratios $z_1$, $\bar{z}_1$ define the position of $x_1$ on the plane $x_1x_2x_3x_4$, while $z_2$, $\bar{z}_2$ define the position of $x_5$ on the plane $x_2x_3x_4x_5$. The angle $\phi$ between the planes is related to $w$ by $w=\sin^2 \frac{\phi}{2}$.

Next we introduce an analog of the radial coordinates~\cite{Hogervorst:2013sma, Costa:2016xah} used for four-point conformal blocks $(r_1, \eta_1, r_2, \eta_2, \hat{w})$:
\begin{equation}
\begin{split}
&z_i=\frac{4 \rho_i}{(1+\rho_i)^2}, \qquad \rho_i = r_i e^{i \theta_i}, \qquad \eta_{i}=\cos \theta_i , \qquad i=1,2, \\
&\hat{w}=\left(\frac{1}{2}-w\right)\sqrt{(1-\eta_1^2)(1-\eta_2^2)}.
\end{split}
\end{equation}
In these coordinates, five-point conformal blocks can be written as an expansion
\begin{equation}\label{blocks-radial}
\begin{split}
&G^{(n_{IJ})}_{(\Delta,l,\Delta',l')}(r_1, \eta_1, r_2, \eta_2, \hat{w})=\\
&\sum_{m_1, m_2=0}^{\infty}r_1^{\Delta+m_1}r_2^{\Delta'+m_2}\sum_{j_1,j_2}\sum_{k=0}^{{\rm min}(j_1,j_2)}c(m_1,m_2,j_1,j_2,k) \hat{w}^{k} C_{j_1-k}^{\left(\nu\right)}\left(\eta _1\right) C_{j_2-k}^{\left(\nu\right)}\left(\eta _2\right),
\end{split}
\end{equation}
where $\nu=\frac{d}{2}-1$ and $C_{n}^{\left(\nu\right)}(x)$ is Gegenbauer polynomial. The parameters $j_1, j_2$ take the values
$$j_1\in [m_1+l, m_1+l-2, m_1+l-4, \ldots , {\rm Mod}(m_1+l, 2) ],     $$ 
$$j_2\in [m_2+l', m_2+l'-2, m_2+l'-4, \ldots , {\rm Mod}(m_2+l', 2) ].    $$

There are two apparent advantages of these coordinates as compared to other representations of the blocks, namely 1) the simplicity of the angular functions and 2) the fact that this expansion only involves two infinite sums over the powers of $r_i$. 

In these coordinates, the two quadratic Casimir equations have the schematic form
\begin{equation}
\left(\sum_{i}R_{i}(r_1, \eta_1, r_2, \eta_2, \hat{w}, \Delta_{12}, \Delta_{3}, \Delta_{45}, \Delta, \Delta', l, l', d)\partial_{i}^{p \leqslant 2} \right)G^{(n_{IJ})}_{(\Delta,l,\Delta',l')} = 0,
\end{equation}
where $\partial_{i}^{p \leqslant 2}$ denotes partial derivatives with respect to the cross-ratios $(r_1, \eta_1, r_2, \eta_2, \hat{w})$ whose order is less than or equal to two. The functions $R_{i}$ are rational functions of their variables.

Now, we can solve the two quadratic Casimir equations order by order in $r_1$ and $r_2$ to compute all of the  coefficients $c(m_1,m_2,j_1,j_2,k)$. These coefficients are rational functions of the conformal dimensions $\Delta_{12}$, $\Delta_{3}$, $\Delta_{45}$, $\Delta$, $\Delta'$, the spins $l$ and $l'$, and the spacetime dimension $d$. We attach a {\fontfamily{lmss}\selectfont Mathematica} notebook with the quadratic Casimir equations where the computation of the coefficients $c(m_1,m_2,j_1,j_2,k)$ is explicitly implemented. We have verified that for small spins the resulting blocks agree with the corresponding ones computed using the recursion relations of~\cite{Poland:2021xjs}.

We observe that at order zero $(m_1=m_2=0)$ there are exactly $1+{\rm min}(l,l')$ linearly independent functions of the angular coordinates $\eta_1$, $\eta_2$, and $\hat{w}$ that satisfy both quadratic Casimir equations. Each of these functions is multiplied by a coefficient $c(0,0,l,l',k)$, where $k=0,1,\ldots {\rm min}(l,l')$. Other coefficients, namely $c(0,0,j_1,j_2,k)$ for $j_1<l$ and $j_2<l'$, are expressed in terms of $c(0,0,l,l',k)$ after solving the Casimir equations at order zero. The choice of conformally-invariant tensor structure in the three-point function  $\langle\mathcal{O}_{\Delta,l}\phi_3\mathcal{O'}_{\Delta',l'} \rangle$ then uniquely fixes the coefficients $c(0,0,l,l',k)$. The quantum number $n_{IJ}$ thus enters the expression on the r.h.s. of eq.~(\ref{blocks-radial}) through the coefficients $c(0,0,l,l',k)$. In the standard box basis, defined in \cite{Costa:2011mg}, these coefficients are given by 
\begin{equation}
c(0,0,l,l',k)= (-1)^{n_{IJ}}\left(-\frac{1}{2}\right)^{k+l+l'}4^{\Delta +\Delta'+2 k}\binom{n_{IJ}}{k}\frac{(1)_{l-k} (1)_{l'-k}}{(\nu )_{l-k} (\nu )_{l'-k}}.
\end{equation}
In this way, we arrive at a unified way to treat all five-point conformal blocks, independent of the spin of the exchanged operators. More importantly, to fix the coefficients of the ansatz (\ref{blocks-radial}),  we do not need to solve quartic Casimir differential equations and the differential equation obtained from the vertex operator as discussed in \cite{Buric:2020dyz}; instead, it is sufficient to just solve two quadratic Casimir differential equations along with the appropriate boundary conditions.

\subsection{Mean-field theory}

As an explicit example, let us consider the five-point correlator $\langle\phi(x_1)\phi(x_2)[\phi,\phi]_{0,0}(x_3)\phi(x_4)\phi(x_5) \rangle$ in mean-field theory in $d>2$ dimensions. We denote the conformal dimension of the field $\phi$ by $\Delta$. The unit-normalized double-twist operators $[\phi,\phi]_{n,l}$ can be schematically represented as 
\begin{equation}\label{mft-oper}
[\phi,\phi]_{n,l}\sim \frac{1}{\sqrt{2}}:\phi \partial_{\mu_1}\ldots\partial_{\mu_l}\partial^{2n} \phi:.
\end{equation}
Their conformal dimensions are given by $\Delta_{n,l}=2\Delta+2n+l$, and their spins by $s_{n,l}=l$. The five-point correlator takes the form
\begin{equation}\label{mft-corr}
\begin{split}
&\langle\phi(x_1)\phi(x_2)[\phi,\phi]_{0,0}(x_3)\phi(x_4)\phi(x_5) \rangle =\left(\frac{x_{24}}{x_{12}x_{23} x_{34} x_{45}}\right)^{2\Delta}\times\\
& \sqrt{2}\left((u_1')^\Delta+ (u_2')^\Delta + (u_1' u_2')^\Delta +\left(\frac{u_1' u_2'}{v_1'}\right)^{\Delta}+\left(\frac{u_1' u_2'}{v_2'}\right)^{\Delta}+\left(\frac{u_1' u_2'}{w'}\right)^{\Delta} \right).
\end{split}
\end{equation}
From eq.~(\ref{blocks-radial}) it follows that at each order in $r_1$ and $r_2$ in the expansion of the five-point correlator there is just a finite number of operators that contribute. Then we can proceed to expand (\ref{mft-corr}) in powers of $r_1$ and $r_2$ to compute the mean-field theory OPE coefficients. 

We denote the products of OPE coefficients that appear in the expansion of the correlator (\ref{mft-corr}) by $P^{n_{IJ}}_{n,l,n',l'}$:
\begin{equation}
P^{n_{IJ}}_{n,l,n',l'}\equiv \lambda_{\phi \phi [\phi, \phi]_{n,l}} \lambda_{\phi \phi [\phi, \phi]_{n',l'}} \lambda^{n_{IJ}}_{[\phi,\phi]_{n,l} [\phi,\phi]_{0,0} [\phi, \phi]_{n',l'}}.
\end{equation}
The OPE coefficients $P^{n_{IJ}}_{n,l,n',l'}$ for the operators with leading twist\footnote{The twist $\tau$ of the operator is defined as the difference between the conformal dimension and the spin, $\tau=\Delta - s$. The operators (\ref{mft-oper}) have twist $\tau_{n,l}=2\Delta+2n$, so that the lowest twist corresponds to $n=0$.} $(n=n'=0)$ were found in \citep{Antunes:2021kmm} and are given by
\begin{equation}
\begin{split}\label{antunes}
P^{n_{IJ}}_{0,l,0,l'}=&\frac{(-1)^{n_{IJ}} 2^{\frac{5}{2}-n_{IJ}} (\Delta )_{\frac{l}{2}} (\Delta )_l (\Delta )_{\frac{l'}{2}} (\Delta )_{l'} \left(l-n_{IJ}+1\right)_{n_{IJ}} \left(l'-n_{IJ}+1\right)_{n_{IJ}}}{l! l'! n_{IJ}! \left(\frac{l-1}{2}+\Delta \right)_{\frac{l}{2}} \left(\frac{l'-1}{2}+\Delta \right)_{\frac{l'}{2}} (\Delta )_{n_{IJ}}}.
\end{split}
\end{equation}
We can expand the correlator (\ref{mft-corr}) in powers of $r_1$ and $r_2$ (in practice we have gone to order $r_1^{2\Delta+m_1} r_2^{2\Delta+m_2}$, with $m_1+m_2\leqslant 6$) and compute the OPE coefficients of all the operators that contribute at this order. The OPE coefficients of operators with leading twist match those given by eq.~(\ref{antunes}). We list here the OPE coefficients of the operators with higher twist that contribute at this order which have not been found from four-point correlators of $\phi$ or $[\phi,\phi]_{0,0}$:
\begin{equation}
\begin{split}
&P^0_{1,0,0,2}=\frac{2 \sqrt{2} \Delta ^3 (\Delta +1) (\Delta +2) (\Delta -\nu )}{(2 \Delta +1) (\nu +1) (2 \Delta -\nu )},\\
&P^0_{1,0,1,0}=\frac{\sqrt{2} \Delta ^3 (\Delta +1) (\Delta -\nu ) (\Delta -\nu +1)}{(\nu +1)^2 (\nu -2 \Delta )^2},\\
&P^2_{0,2,1,2}=\frac{\sqrt{2} \Delta ^3 (\Delta +1)^2 (\Delta +2) (\Delta -\nu )}{(4 \Delta  (\Delta +2)+3) (\nu +3) (2 \Delta -\nu +2)},\\
&P^1_{0,2,1,2}=-\frac{4 \sqrt{2} \Delta ^3 (\Delta +1)^2 (\Delta +2)^2 (\Delta -\nu )}{(4 \Delta  (\Delta +2)+3) (\nu +3) (2 \Delta -\nu +2)},\\
&P^0_{0,2,1,2}=\frac{2 \sqrt{2} \Delta ^3 (\Delta +1)^3 (\Delta +2)^2 (\Delta -\nu )}{(4 \Delta  (\Delta +2)+3) (\nu +3) (2 \Delta -\nu +2)},
\end{split}
\end{equation}

\begin{equation}
\begin{split}
&P^0_{1,0,1,2}=\frac{\sqrt{2} \Delta ^3 (\Delta +1)^2 (\Delta +2) (\Delta +3) (\Delta -\nu ) (\Delta -\nu +1)}{(2 \Delta +3) (\nu +1) (\nu +3) (2 \Delta -\nu ) (2 \Delta -\nu +2)},\\
&P^0_{1,0,2,0}=\frac{\Delta ^3 (\Delta +1)^2 (\Delta +2) (\Delta -\nu )^2 (\Delta -\nu +1) (\Delta -\nu +2)}{\sqrt{2} (\nu +1)^2 (\nu +2) (2 \Delta -2 \nu +1) (2 \Delta -\nu ) (2 \Delta -\nu +1) (2 \Delta -\nu +2)}.
\end{split}
\end{equation}
These OPE coefficients can be obtained from the following generalization of eq.~(\ref{antunes}):
\begin{equation}\label{mft-full-ope}
\begin{split}
P^{n_{IJ}}_{n,l,n',l'}=&\frac{(-1)^{n_{IJ}} 2^{\frac{5}{2}-n_{IJ}} (l-n_{IJ}+1)_{n_{IJ}} (l'-n_{IJ}+1)_{n_{IJ}} (\Delta )_{\frac{l}{2}+n}(\Delta )_{\frac{l'}{2}+n'}}{l! l'! n! n'! n_{IJ}! (l+\nu +1) (l'+\nu +1) (l+\nu +2)_{n-1}(l'+\nu +2)_{n'-1}  }\\
\times &\frac{(\Delta -\nu )_n (\Delta -\nu )_{n'}(\Delta -\nu )_{n+n'}}{\left(\frac{l-1}{2}+n+\Delta \right)_{\frac{l}{2}}\left(\frac{l'-1}{2}+n'+\Delta \right)_{\frac{l'}{2}} (n+2 \Delta -2 \nu -1)_n (n'+2 \Delta -2 \nu -1)_{n'} }\\
\times &\frac{(\Delta )_{l+n+n'} (\Delta )_{l'+n+n'}}{(l+n+2 \Delta -\nu -1)_n (l'+n'+2 \Delta -\nu -1)_{n'} (\Delta )_{n+n'+n_{IJ}}}.
\end{split}
\end{equation}
We conjecture that this is the correct formula for all $n$, $n'$, $l$, $l'$, $n_{IJ}$.

\section{OPE truncation in five-point correlators}
\label{sec:truncation}

In this section we apply our knowledge of the five-point conformal blocks to both the free theory and the critical Ising model in $d=3$ to approximately compute OPE coefficients involving one scalar and two spinning operators. We truncate the OPEs in the five-point correlators by including just a finite set of exchanged operators with conformal dimensions below a certain cutoff $\Delta\leqslant \Delta_{\rm cutoff}$. We then determine the OPE coefficients by demanding that the truncated correlators approximately satisfy the crossing relation.

We consider five-point correlation functions of scalar operators $\langle\phi_1(x_1)\phi_1(x_2)\phi_2(x_3)\phi_1(x_4)\phi_1(x_5)\rangle$. The correlator can be expanded in terms of conformal blocks as
\begin{equation}\label{first-channel}
\begin{split}
&\langle\phi_1(x_1)\phi_1(x_2)\phi_2(x_3)\phi_1(x_4)\phi_1(x_5)\rangle =\\
&\sum_{(\mathcal{O}_{\Delta,l},\mathcal{O'}_{\Delta',l'})}\sum_{n_{IJ}=0}^{{\rm min}(l,l')} \lambda_{\phi_1\phi_1 \mathcal{O}_{\Delta,l}}\lambda_{\phi_1\phi_1 \mathcal{O'}_{\Delta',l'}}\lambda_{\mathcal{O}_{\Delta,l} \phi_2 \mathcal{O'}_{\Delta',l'}}^{n_{IJ}}P(x_i)G^{(n_{IJ})}_{(\Delta, l, \Delta', l')}(u_1',v_1',u_2',v_2',w'),
\end{split}
\end{equation}
where $\Delta_{1}$ and $\Delta_{2}$ are conformal dimensions of $\phi_1$ and $\phi_2$, respectively. Here, we use the standard box basis for the independent conformally-invariant tensor structures in the three-point functions $\langle\mathcal{O}_{\Delta,l}\phi_3 \mathcal{O}_{\Delta',l'} \rangle$, as in \cite{Poland:2021xjs}.

The same five-point correlation function can be written in a different channel by considering the operator product expansions $\phi_1(x_1)\times \phi_1(x_4)$ and $\phi_1(x_2)\times \phi_1(x_5)$. This corresponds to exchanging $x_2 \leftrightarrow x_4$ in eq.~(\ref{first-channel}). It is easy to check that under this exchange, the cross-ratios behave as
\begin{equation}\label{exchange-cross-ratios}
\begin{split}
&u_1'\leftrightarrow v_1',\\
&u_2'\leftrightarrow v_2'.
\end{split}
\end{equation}
Now, the five-point correlation function is given by
\begin{equation}\label{second-channel}
\begin{split}
&\langle\phi_1(x_1)\phi_1(x_2)\phi_2(x_3)\phi_1(x_4)\phi_1(x_5)\rangle =\\
&\sum_{(\mathcal{O}_{\Delta,l},\mathcal{O'}_{\Delta',l'})}\sum_{n_{IJ}=0}^{{\rm min}(l,l')} \lambda_{\phi_1\phi_1 \mathcal{O}_{\Delta,l}}\lambda_{\phi_1\phi_1 \mathcal{O'}_{\Delta',l'}}\lambda_{\mathcal{O}_{\Delta,l} \phi_2 \mathcal{O'}_{\Delta',l'}}^{n_{IJ}}\tilde{P}(x_i)G^{(n_{IJ})}_{(\Delta, l, \Delta', l')}(v_1',u_1',v_2',u_2',w'),
\end{split}
\end{equation}
where 
\begin{equation}
\tilde{P}(x_i)=\frac{1}{x_{14}^{2\Delta_{1}} x_{23}^{\Delta_{2}} x_{25}^{2\Delta_{1}}}\left(\frac{x_{24}}{x_{34}}\right)^{\Delta_{2}}.
\end{equation}

We will expand around the following configuration of operators:
\begin{equation}
x_3\to \infty, \quad x_{12}^2=1, \quad x_{14}^2=1, \quad x_{24}^2=1, \quad x_{25}^2=1, \quad x_{45}^2=1, \quad x_{15}^2=\frac{3}{2}.
\end{equation}
In terms of cross-ratios, this choice corresponds to
\begin{equation}\label{configuration}
u_1'=v_1'=u_2'=v_2'=1, \qquad w'=\frac{3}{2},
\end{equation}
or in terms of the radial coordinates to
\begin{equation}
r_1=r_2=2-\sqrt{3}, \qquad \eta_1=\eta_2=\hat{w}=0.
\end{equation}
This configuration is convenient as it is symmetric with respect to $u'_{i}\leftrightarrow v'_{i}$ exchange; moreover, it  is particularly easy to numerically compute conformal blocks here since they converge quickly in this region.

The associativity of the operator product expansion requires that eq.~(\ref{first-channel}) is equal to eq.~(\ref{second-channel}), and by matching these we get
\begin{equation}\label{bootstrap-eq-one}
\begin{split}
&\sum_{(\mathcal{O}_{\Delta,l},\mathcal{O'}_{\Delta',l'})}\sum_{n_{IJ}=0}^{{\rm min}(l,l')} \lambda_{\phi_1\phi_1 \mathcal{O}_{\Delta,l}}\lambda_{\phi_1\phi_1 \mathcal{O'}_{\Delta',l'}} \lambda_{\mathcal{O}_{\Delta,l} \phi_2 \mathcal{O'}_{\Delta',l'}}^{n_{IJ}}\times\\ 
&\Big((v_1' v_2')^{\Delta_{1}} G^{(n_{IJ})}_{(\Delta, l, \Delta', l')}(u_1',v_1',u_2',v_2',w')- (u_1' u_2')^{\Delta_{1}}G^{(n_{IJ})}_{(\Delta, l, \Delta', l')}(v_1',u_1',v_2',u_2',w')\Big)=0.
\end{split}
\end{equation}
Next, we separate the identity operator that appears in the two pairs of contributions: $(\hat{1},\phi_2)$ and $(\phi_2,\hat{1})$. The corresponding five-point conformal blocks are given by 
\begin{equation}
\begin{split}
G^{(0)}_{(0, 0, \Delta_{2}, 0)}(u_1',v_1',u_2',v_2',w')={u_2'}^{\Delta_{2}/2},\\
G^{(0)}_{(\Delta_{2}, 0, 0, 0)}(u_1',v_1',u_2',v_2',w')={u_1'}^{\Delta_{2}/2}.
\end{split}
\end{equation}
With this, we write eq.~(\ref{bootstrap-eq-one}) as
\begin{equation}\label{bootstrap-eq}
\sum_{\mathcal{O}_{\Delta,l}\neq \hat{1}}\sum_{\mathcal{O'}_{\Delta',l'}\neq \hat{1}}\sum_{n_{IJ}=0}^{{\rm min}(l,l')} \frac{\lambda_{\phi_1\phi_1 \mathcal{O}_{\Delta,l}}\lambda_{\phi_1\phi_1 \mathcal{O'}_{\Delta',l'}} \lambda_{\mathcal{O}_{\Delta,l} \phi_2 \mathcal{O'}_{\Delta',l'}}^{n_{IJ}}}{\lambda_{\phi_1 \phi_1 \phi_2}}\mathcal{F}^{n_{IJ}}_{\Delta, l, \Delta', l'}(u_1',v_1',u_2',v_2',w')-1  =0,
\end{equation}
where
\begin{equation}\label{Fdef}
\begin{split}
&\mathcal{F}^{n_{IJ}}_{\Delta, l, \Delta', l'}(u_1',v_1',u_2',v_2',w')=\\
&\left(\frac{(v_1' v_2')^{\Delta_{1}}G^{(n_{IJ})}_{(\Delta, l, \Delta', l')}(u_1',v_1',u_2',v_2',w')-(u_1' u_2')^{\Delta_{1}}G^{(n_{IJ})}_{(\Delta, l, \Delta', l')}(v_1',u_1',v_2',u_2',w')}{(u_1' u_2')^{\Delta_{1}}({v_1'}^{\Delta_{2}/2}+{v_2'}^{\Delta_{2}/2})-(v_1' v_2')^{\Delta_{1}}({u_1'}^{\Delta_{2}/2}+{u_2'}^{\Delta_{2}/2})}\right).
\end{split}
\end{equation}
We assume $\lambda_{\phi_1 \phi_1 \phi_2}\neq 0$, which implies that the contribution of the exchanged identity operator is non-vanishing. If this was not the case, one could still run the same algorithm without including the identity contribution.
For convenience, we use the following parametrization of the cross-ratios:
\begin{equation}\label{new-coordinates}
\begin{split}
u_1'=& \frac{1}{4} \left((a^{-}+a^{+})^2-b^{-}-b^{+}\right) ,\\
v_1'=& \frac{1}{4} \left((a^{-}+a^{+}-2)^2-b^{-}-b^{+}\right),\\
u_2'=& \frac{1}{4} \left((a^{+}-a^{-})^2+b^{-}-b^{+}\right) ,\\
v_2'=& \frac{1}{4} \left((a^{+}-a^{-}-2)^2+b^{-}-b^{+}\right),\\
w' =& \frac{1}{4} \Big((a^{-}+a^{+})^2+2 (a^{+}-a^{-}-2) (a^{-}+a^{+})+(a^{+}-a^{-}-4) (a^{+}-a^{-})\\
&+2 (2 w-1) \sqrt{b^{+}-b^{-}} \sqrt{b^{-}+b^{+}}-2 b^{+}+4\Big).
\end{split}
\end{equation}
In these coordinates, the configuration (\ref{configuration}) is given by
\begin{equation}\label{configurationab}
a^{+}=1,\qquad b^{+}=-3, \qquad a^{-}=b^{-}=0, \qquad w=\frac{1}{2}.
\end{equation}
We next apply a modification of the method developed in \cite{Gliozzi:2013ysa, Gliozzi:2014jsa, Li:2017ukc} to solve eq.~(\ref{bootstrap-eq}). Namely, we include a finite number of contributions $\mathcal{O}_{\Delta,l}$ and $\mathcal{O'}_{\Delta',l'}$ in the OPEs, thus truncating the operator product expansions. In our calculation, we input all previously determined CFT data from the theory of interest, with the exception of $\Delta_{1}$. Ordinarily, this data is obtained  from the conformal bootstrap of four-point correlators.  One reason that it is advantageous to input all other conformal dimensions is the fact that the five-point conformal blocks are very cumbersome in their generic form, which makes the evaluation of the generic blocks and the computation of the spectrum of the theory difficult. However, we can easily keep $\Delta_{1}$ generic since the conformal blocks do not depend on it and it serves as a sanity check for our calculation. We compare the value we obtain for $\Delta_1$ by our method to the values obtained by the conformal bootstrap of four-point correlators. 

Now, in order to generate more independent equations for the unknown OPE coefficients and conformal dimension $\Delta_{1}$, we take derivatives of the truncated eq.~(\ref{bootstrap-eq}) with respect to the coordinates $(a^{\pm},b^{\pm},w)$.  Crucially, we generate more independent equations than the number of unknown OPE coefficients plus one (one is for $\Delta_{1}$, which we also treat as an unknown). Then, we obtain an over-determined system of equations that is not possible to solve exactly. Instead, we sum up the squares of each of these equations and find the minimum of the resulting cost function.\footnote{Let us remark that truncation approaches to the four-point bootstrap which minimize a cost function using stochastic methods have been developed recently in~\cite{Kantor:2021kbx, Kantor:2021jpz,  Laio:2022ayq, Kantor:2022epi}. It would be interesting to explore the use of stochastic optimizers in the five-point bootstrap, but we defer this to future work.}

The equations obtained by taking the derivatives of eq.~(\ref{bootstrap-eq}) are denoted by $e_i(\Delta_{1},\lambda)$
\begin{equation}\label{equations}
\begin{split}
& e_{i}(\Delta_{1},\lambda)=\\
& \mathcal{D}_{i}\left( \sum_{\mathcal{O}_{\Delta,l},\mathcal{O'}_{\Delta',l'}}^{{\rm trunc.}}\sum_{n_{IJ}=0}^{{\rm min}(l,l')} \frac{\lambda_{\phi_1\phi_1 \mathcal{O}_{\Delta,l}}\lambda_{\phi_1\phi_1 \mathcal{O'}_{\Delta',l'}} \lambda_{\mathcal{O}_{\Delta,l} \phi_2 \mathcal{O'}_{\Delta',l'}}^{n_{IJ}}}{\lambda_{\phi_1 \phi_1 \phi_2}}\mathcal{F}^{n_{IJ}}_{\Delta, l, \Delta', l'}(a^{\pm},b^{\pm},w)-1   \right)\Bigg|_{(\ref{configurationab})},
\end{split}
\end{equation}
where $i\geqslant1$ and $\mathcal{D}_{i}$ represent a choice of derivatives, a subset of all derivatives up to  third order, which we select such that the value of $\Delta_{1}$ that we find in this bootstrap algorithm lies closest to the actual value known from the bootstrap of the four-point correlators. The label $e_{0}$ represents the equation that we obtained  by not taking any derivatives of eq.~(\ref{bootstrap-eq}), $\mathcal{D}_{0}\equiv 1$. Lastly, $\lambda$ represents all unknown OPE coefficients. 

Let us now define the cost functions
\begin{equation}
f_{\{r_i\}}(\Delta_{1},\lambda)= \sum_{i=0}^{{\rm dim\,}\mathcal{D}}r_i\left(\frac{e_{i}(\Delta_{1},\lambda)}{e_{i}(\Delta_{1},0)}\right)^2,
\end{equation}
where $r_{i}$ are (pseudo-)randomly generated real numbers, $r_{i}\in [0,1]$, which assign random weights to each term. We proceed as follows. For each set of randomly generated numbers $r_i$, we find the minimum of the function $f_{\{r_i\}}(\Delta_{1},\lambda)$ in order to compute $\Delta_{1}$ and the unknown OPE coefficients $\lambda$. We then repeat this procedure for different choices of ${\rm dim\,}\mathcal{D}$. For each choice of ${\rm dim\,}\mathcal{D}$ we choose the set of derivatives of the equations $e_i(\Delta_1,\lambda)$ such that $\Delta_{1}$ lies closest to the exact value with the minimal standard deviation. Finally, we combine the data obtained with different values of ${\rm dim\,}\mathcal{D}$ into a single set. Subsequently, we take the average of the values for $\Delta_1$ and unknown OPE coefficients and use their standard deviations as a rough estimate of the error bars. 

\subsection{Five-point correlator $\langle\phi\phi \phi^2 \phi\phi\rangle$ in the free theory}

As a testing ground, we first consider the five-point function $\langle\phi(x_1)\phi(x_2) \phi^2(x_3) \phi(x_4)\phi(x_5)\rangle$ in the free scalar theory in $d=3$. We truncate the OPEs and include in eq.~(\ref{bootstrap-eq}) all exchanges containing the operators $\hat{1}$, $\phi^2 \equiv [\phi,\phi]_{0,0}$, $[\phi,\phi]_{0,2}$, and $[\phi,\phi]_{0,4}$. 
The operator $[\phi,\phi]_{0,2}$ is the stress tensor of the free theory $[\phi,\phi]_{0,2}\equiv T_{\mu\nu}$ and $[\phi,\phi]_{0,4}$ is the spin-4 conserved current $[\phi,\phi]_{0,4}\equiv C_{\mu\nu\rho\sigma}$. 

We take $\Delta_{\phi}$ to be the conformal dimension of the field $\phi(x)$ and treat it as an unknown as well as the OPE coefficients $\lambda_{T\phi^2 T}^{0}$, $\lambda_{T\phi^2 C}^{0}$, and $\lambda_{C\phi^2 C}^{n_{IJ}}$.\footnote{The conservation of the stress tensor allows us to write $\lambda_{T\phi^2 T}^{2}$ and $\lambda_{T\phi^2 T}^{1}$ in terms of  $\lambda_{T\phi^2 T}^{0}$. See, for example, eq.~(4.123) in \citep{Poland:2021xjs}. The same is true for a general spinning operator like $[\phi,\phi]_{0,l}$: $\lambda_{T \phi^2 [\phi,\phi]_{0,l}}^{1}$ and $\lambda_{T \phi^2 [\phi,\phi]_{0,l}}^{2}$ can be written in terms of $\lambda_{T \phi^2 [\phi,\phi]_{0,l}}^{0}$. See eqs.~(B.4) and (B.5) in \citep{Meltzer:2018tnm}. 
} The rest of the conformal dimensions and OPE coefficients are fixed to their exact values: 
\begin{equation}
\begin{split}
&\Delta_{[\phi,\phi]_{0,l}} =1+2l,\\
&\lambda_{\phi\phi [\phi,\phi]_{0,l}}=(1+(-1)^l)\left(\frac{\left(\frac{1}{2}\right)_{\frac{l}{2}} \left(\frac{1}{2}\right)_l}{2l!\left(\frac{l}{2}\right)_{\frac{l}{2}}}\right)^{\frac{1}{2}},\, \lambda_{\phi^2\phi^2 [\phi,\phi]_{0,l}}=(1+(-1)^l)\left(\frac{2 \left(\frac{1}{2}\right)_{\frac{l}{2}}\left(\frac{1}{2}\right)_l}{l! \left(\frac{l}{2}\right)_{\frac{l}{2}}}\right)^{\frac{1}{2}}.
\end{split}
\end{equation}
We remark that the cutoff conformal dimension in this case is $\Delta_{\rm cutoff}=5$. We scan over all choices of equations $e_i$ obtained by taking up to three derivatives of the eq.~(\ref{bootstrap-eq}) and for each value of ${\rm dim\,}\mathcal{D}$ we pick the set $\{e_i|i=0,1,2,\ldots {\rm dim\,}\mathcal{D}\}$ that gives the value of $\Delta_{\phi}$ closest to $1/2$ with the minimal standard deviation. The sets of derivatives that we use are given by eq.~(\ref{ftdtwo}).  Upon taking the average of the data obtained for different values of ${\rm dim\,}\mathcal{D}$,  we ultimately arrive at the results given in table~\ref{freetheorytable}.
\begin{table}[H]\center
\begin{tabular}{l|l|l|}
\cline{2-3}
                                                & truncation   & exact     \\ \hline
\multicolumn{1}{|l|}{$\Delta_{\phi}$}           & 0.5000(3)    & 0.500000  \\ \hline
\multicolumn{1}{|l|}{$\lambda^{0}_{T\phi^2 T}$} & 0.52(1)      & 0.530330  \\
\multicolumn{1}{|l|}{$\lambda^{0}_{T\phi^2 C}$} & 0.21(2)      & 0.226428  \\
\multicolumn{1}{|l|}{$\lambda^{4}_{C\phi^2 C}$} & 0.03(2)     & 0.022097  \\ \hline
\multicolumn{1}{|l|}{$\lambda^{3}_{C\phi^2 C}$} & -1.0(3)      & -0.618718 \\
\multicolumn{1}{|l|}{$\lambda^{2}_{C\phi^2 C}$} & 5.1(9)       & 2.320194  \\
\multicolumn{1}{|l|}{$\lambda^{1}_{C\phi^2 C}$} & 2(2)        & -1.546796 \\
\multicolumn{1}{|l|}{$\lambda^{0}_{C\phi^2 C}$} & 1.2(5)        & 0.096675  \\ \hline
\end{tabular}
\caption{$\Delta_{\phi}$ and OPE coefficients that we treat as unknown in the free theory found by minimizing the cost functions and averaging over the obtained values for different sets of randomly generated numbers $r_i$ and different choices of ${\rm dim\,}\mathcal{D}$.
The right column shows the exact results in the free theory, computed in \cite{Antunes:2021kmm}. In appendix~\ref{app:exact} we give the exact expressions for these OPE coefficients. The OPE coefficients in the bottom part of the table have standard deviations significantly larger than the OPE coefficients in the top part of the table. We are not able to compute these accurately using this truncation.}\label{freetheorytable}
\end{table}

From the table, we readily see that the truncated five-point conformal bootstrap is quite successful at computing the OPE coefficients.  We note that although we are able to determine most of the error bars accurately, a few of them are underestimated. This suggests that the given method for error estimation is not rigorous. To estimate the error bars more precisely one would need to  better approximate the truncated part of the correlator. We also note  that the error bars of the OPE coefficients $\lambda^{n_{IJ}}_{C\phi^2 C}$, $n_{IJ}=0,1,2,3$, in table~\ref{freetheorytable} are significantly larger then  those for the OPE coefficients $\lambda^{0}_{T\phi^2 T}$, $\lambda^{0}_{T\phi^2 C}$ and $\lambda^{4}_{C\phi^2 C}$. The reason for this is the fact that these operators and structures contribute to the crossing relation at the same order of magnitude as the truncated part of the correlator; hence, it is not surprising that we are not able to reliably determine their OPE coefficients.  


\begin{figure}[H]
\centering
\includegraphics[width=0.38\textwidth]{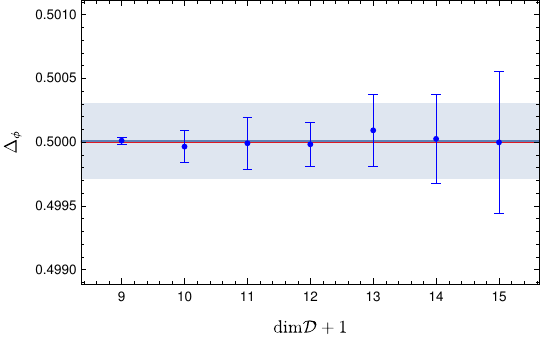}
\hspace{0.05\textwidth}
\includegraphics[width=0.38\textwidth]{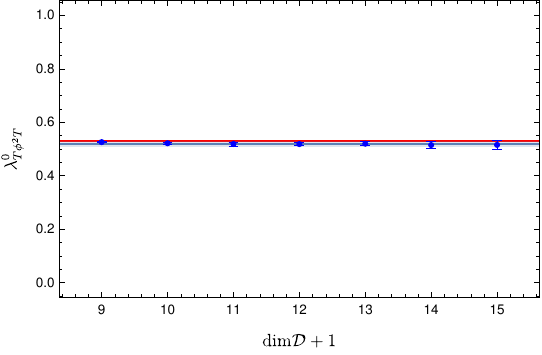}\\
\includegraphics[width=0.38\textwidth]{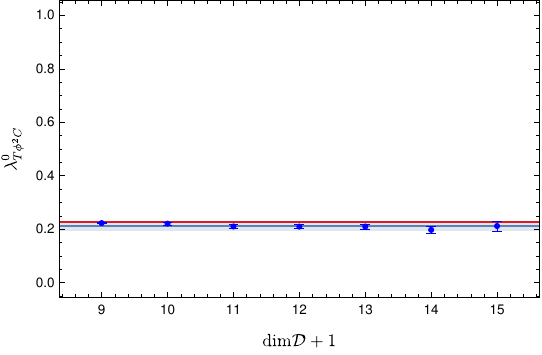}
\hspace{0.05\textwidth}
\includegraphics[width=0.38\textwidth]{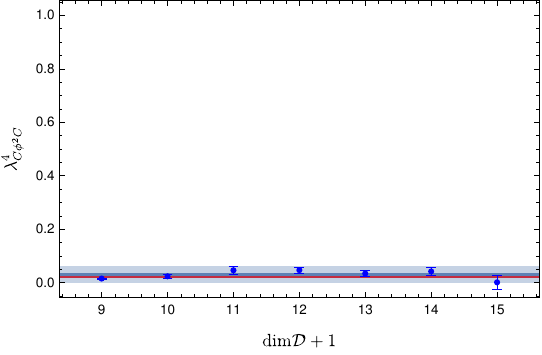}\\
\includegraphics[width=0.38\textwidth]{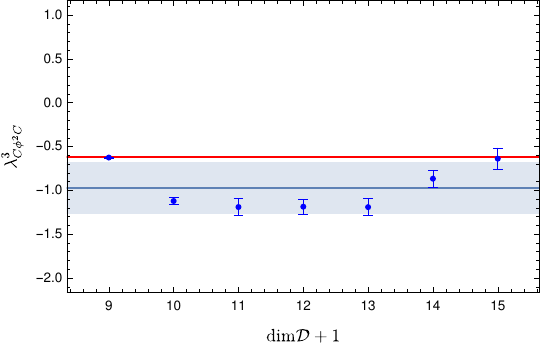}
\hspace{0.05\textwidth}
\includegraphics[width=0.38\textwidth]{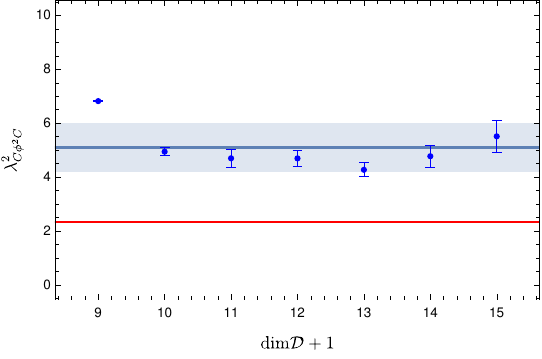}\\
\includegraphics[width=0.38\textwidth]{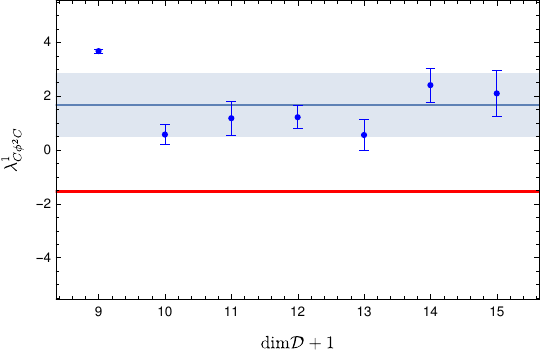}
\hspace{0.05\textwidth}
\includegraphics[width=0.38\textwidth]{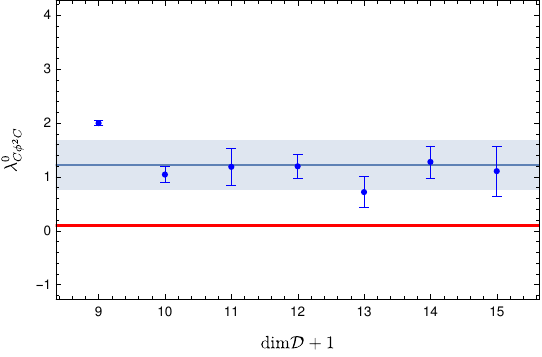}\\
\caption{{\small $\Delta_{\phi}$ and the unknown OPE coefficients in the 3d free scalar theory computed by minimizing the cost functions $f_{\{r_i\}}(\Delta_{\phi},\lambda)$ defined for different choices of ${\rm dim\,}\mathcal{D}$. For each choice of ${\rm dim\,}\mathcal{D}$, the set of derivatives that we use is chosen such that $\Delta_{\phi}$ is closest to $1/2$ with the minimal standard deviation after averaging over data obtained by randomly selecting the weights $r_i$ in the cost function. Horizontal red lines represent the exact data of the free theory. Horizontal blue lines are the mean values of the OPE coefficients given in table \ref{freetheorytable} and the blue strips are the error bars. One can observe that the standard deviations of $\Delta_{\phi}$ increase as we increase ${\rm dim\,}\mathcal{D}$. Also, it is obvious that the OPE coefficients  we obtain have a systematic error due to the truncation of the OPE. Still, the systematic errors in the coefficients $\lambda_{T\phi^2 T}^0$, $\lambda_{T\phi^2 C}^0$, and $\lambda_{C\phi^2 C}^4$ are relatively small and  are accurately estimated within the standard deviations. Other OPE coefficients have larger systematic errors. We conclude that we are unable to accurately compute them using this truncation.}}
\label{free-theory-plots}
\end{figure}

We can also exploit the fact that the spin-4 operator is a conserved current in the free theory. Therefore, there are Ward identities relating the OPE coefficients $\lambda_{C\phi^2 C}^{n_{IJ}}$.  In particular, these Ward identity relations are given by
\begin{equation}\label{Ward-C}
\lambda_{C\phi^2 C}^{3}=-28 \lambda_{C\phi^2 C}^{4}, \quad \lambda_{C\phi^2 C}^{2}=105 \lambda_{C\phi^2 C}^{4}, \quad \lambda_{C\phi^2 C}^{1}=-70\lambda_{C\phi^2 C}^{4}, \quad \lambda_{C\phi^2 C}^{0}=\frac{35}{8} \lambda_{C\phi^2 C}^{4}.
\end{equation}
Now, we can minimize the cost functions to compute $\lambda^{0}_{T\phi^2 T}$, $\lambda^{0}_{T\phi^2 C}$, and $\lambda^{4}_{C\phi^2 C}$. The set of derivatives that we use is given by eq.~(\ref{ftd}). Again, these sets are chosen such that $\Delta_{\phi}$ is closest to the exact value with the minimal standard deviation.  Here, we scan over the set of equations~(\ref{equations}) obtained by taking three or fewer derivatives with respect to the cross-ratios $(a^{\pm},b^{\pm},w)$. The results are given in table~\ref{tabletwo}.
\begin{table}[H]\center
\begin{tabular}{l|l|l|}
\cline{2-3}
                                                & truncation   & exact     \\ \hline
\multicolumn{1}{|l|}{$\Delta_{\phi}$}           & 0.4999(6)    & 0.500000  \\ \hline
\multicolumn{1}{|l|}{$\lambda^{0}_{T\phi^2 T}$} & 0.513(9)      & 0.530330  \\
\multicolumn{1}{|l|}{$\lambda^{0}_{T\phi^2 C}$} & 0.21(2)      & 0.226428  \\
\multicolumn{1}{|l|}{$\lambda^{4}_{C\phi^2 C}$} & 0.02(2)     & 0.022097  \\ \hline
\end{tabular}
\caption{$\Delta_{\phi}$ and OPE coefficients that we treat as unknown in the free theory found by minimizing the cost functions and averaging over the obtained values for different sets of randomly generated numbers $r_i$ and different values of ${\rm dim\,}\mathcal{D}$.
The third column shows exact results in the free theory, computed in \cite{Antunes:2021kmm}. The OPE coefficients $\lambda_{C\phi^2 C}^{n_{IJ}}$, for $n_{IJ}=0,1,2,3$, are related to $\lambda^{4}_{C\phi^2 C}$ through the Ward identities given by eq.~(\ref{Ward-C}).}
\label{tabletwo}
\end{table}
It is evident that $\lambda_{C \phi^2 C}^4$ is now closer to its exact value, with a smaller error. On the other hand, $\lambda_{T \phi^2 T}$ is now a little farther away. In all cases the exact values are still within two standard deviations as estimated by our error bars, lending some reassurance that this method of error estimation can be used for other theories for which exact solutions are not known.

\subsection{Five-point correlator $\langle\sigma\sigma\epsilon\sigma\sigma\rangle$ in the critical 3d Ising model}

Next we consider the five-point correlation function $\langle\sigma(x_1)\sigma(x_2)\epsilon(x_3)\sigma(x_4)\sigma(x_5)\rangle$ in the critical 3d Ising model. We choose to truncate the $\sigma \times \sigma$ OPEs and include the operators $\hat{1}$, $\epsilon$, $\epsilon'$, $T_{\mu\nu}$, and $C_{\mu\nu\rho\sigma}$.  In particular,  in eq.~(\ref{bootstrap-eq}) we include  all  possible pairs of these operators.

The unknown OPE coefficients in eq.~(\ref{bootstrap-eq}) consist of the set $\{\lambda_{T\epsilon T}^{0}$, $\lambda_{T\epsilon C}^{0}$, $\lambda_{\epsilon' \epsilon C}$, $\lambda_{\epsilon' \epsilon \epsilon'}$, $\lambda_{C\epsilon C}^{n_{IJ}}\}$. All other OPE coefficients that appear in eq.~(\ref{bootstrap-eq}), when it is truncated to include only the given operators, have been previously computed in \cite{Simmons-Duffin:2016wlq}.\footnote{Our normalization of conformal blocks differs from the one in \cite{Simmons-Duffin:2016wlq} by the factor $2^{l/2}$. Our OPE coefficients $\lambda$ are given by $\lambda_{\sigma \sigma \mathcal{O}_{\Delta,l}}=2^{l/2} f_{\sigma \sigma \mathcal{O}_{\Delta,l}}$ and $\lambda_{\epsilon \epsilon \mathcal{O}_{\Delta,l}}=2^{l/2} f_{\epsilon \epsilon \mathcal{O}_{\Delta,l}}$, where the label  $f$ denotes the OPE coefficients in \cite{Simmons-Duffin:2016wlq}.} In our study we fix  the following conformal dimensions and OPE coefficients:
\begin{equation}
\begin{split}
&\Delta_\epsilon = 1.412625, \quad \Delta_{\epsilon'} = 3.82968, \quad \Delta_C = 5.022665,\\
&\lambda_{\sigma\sigma\epsilon}=1.0518537,\quad \lambda_{\epsilon\epsilon\epsilon}=1.532435, \quad \lambda_{\sigma\sigma\epsilon'}=0.053012,\quad \lambda_{\epsilon\epsilon\epsilon'}=1.5360,\\
&\lambda_{\sigma\sigma T}=0.65227552,\quad \lambda_{\epsilon\epsilon T}= 1.7782942, \quad \lambda_{\sigma\sigma C}=0.276304, \quad \lambda_{\epsilon\epsilon C}=0.99168.
\end{split}
\end{equation}
We already know from the $\varepsilon$-expansion and the four-point bootstrap that the above operators are not
degenerate (see e.g.~\cite{Henriksson:2022rnm}), but in case they were, our method would be sensitive only
to the sums of the OPE coefficients of any degenerate operators.
In our case, the cutoff conformal dimension is given by $\Delta_{\rm cutoff}\approx 5.023$. To increase the cutoff value, one would have to also include contributions of the spin-6 operator, which we have not yet pursued due to the long conformal block computation time in our current implementation.  Again, for each value of ${\rm dim\,}\mathcal{D}$, we select a set of equations $\{e_i(\Delta_{\sigma},\lambda) | i=0,1,\ldots, {\rm dim\,}\mathcal{D}\}$ that gives $\Delta_{\sigma}$ closest to the exact value with the minimal standard deviation. The sets of derivatives that we now use are given by eq.~(\ref{imd}).

\begin{figure}[t!]
\centering
\includegraphics[width=0.40\textwidth]{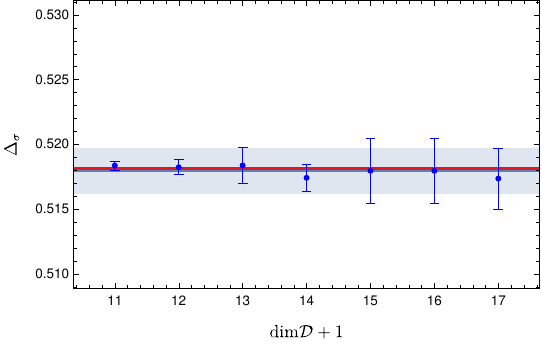}
\hspace{0.05\textwidth}
\includegraphics[width=0.40\textwidth]{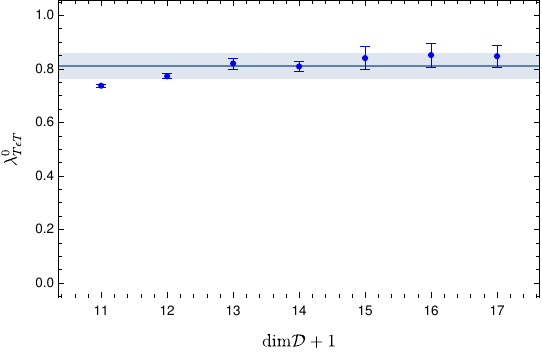}
\includegraphics[width=0.40\textwidth]{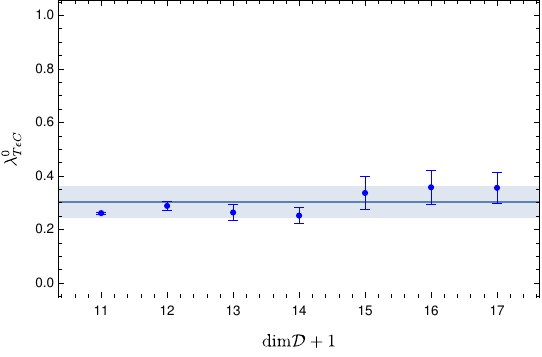}
\hspace{0.05\textwidth}
\includegraphics[width=0.40\textwidth]{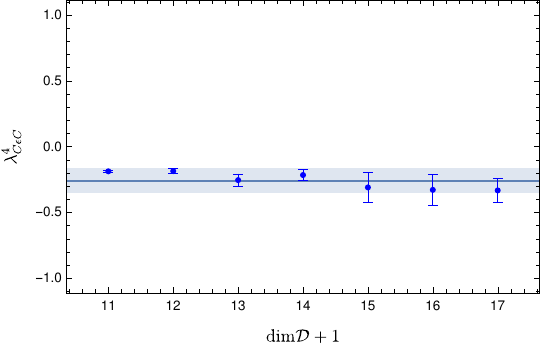}\\
\caption{{\small $\Delta_{\sigma}$ and the unknown OPE coefficients in the critical 3d Ising CFT computed by minimizing the cost functions $f_{\{r_i\}}(\Delta_{\phi},\lambda)$ defined with different choices of ${\rm dim\,}\mathcal{D}$. For each choice of ${\rm dim\,}\mathcal{D}$, the set of derivatives that we use is chosen such that $\Delta_{\sigma}$ is closest to the exact value with the minimal standard deviation after averaging over data obtained by randomly selecting the weights $r_i$ in the cost function. The horizontal red line represents the exact value of $\Delta_{\sigma}$. The horizontal blue lines are the mean values of the data and the blue strips represent the error bars given in table~\ref{3dIsingtable}.   One can observe that the standard deviations increase as we increase ${\rm dim\,}\mathcal{D}$. 
Here one can notice that the values of OPE coefficients $\lambda_{T\epsilon T}^0$, $\lambda_{T\epsilon C}^0$, and $\lambda_{C\epsilon C}^4$ are not very sensitive to the choice of derivatives in the cost function and consequently the standard deviations for these coefficients after averaging the data for all values of ${\rm dim\,}\mathcal{D}$ shown in the plots are small.}}
\label{Ising-theory-plots-1}
\end{figure}

\begin{figure}[t!]
\centering
\includegraphics[width=0.40\textwidth]{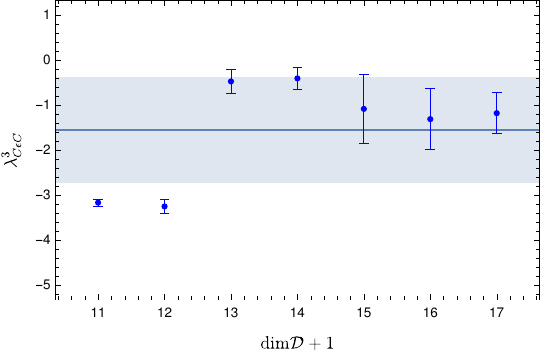}
\hspace{0.05\textwidth}
\includegraphics[width=0.40\textwidth]{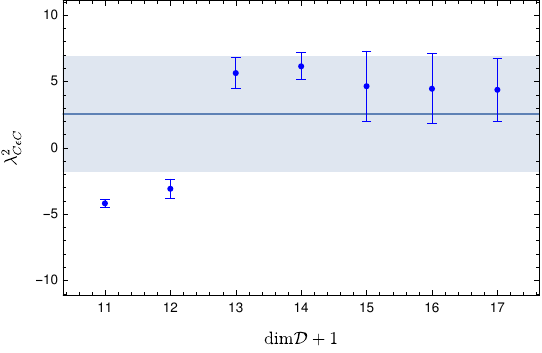}
\includegraphics[width=0.40\textwidth]{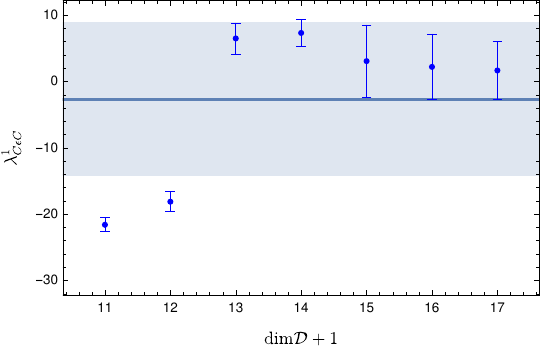}
\hspace{0.05\textwidth}
\includegraphics[width=0.40\textwidth]{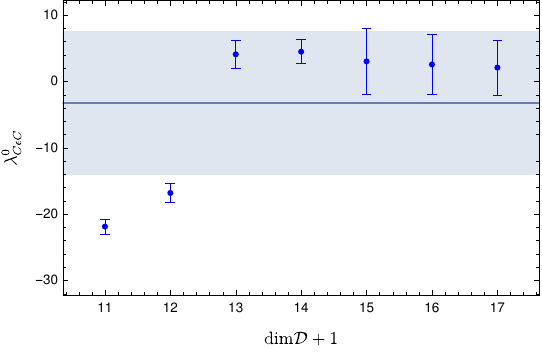}
\includegraphics[width=0.40\textwidth]{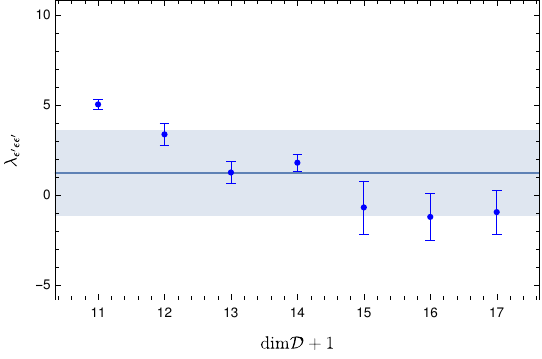}
\hspace{0.05\textwidth}
\includegraphics[width=0.40\textwidth]{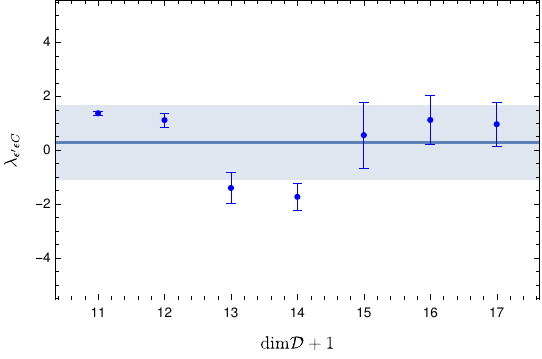}\\
\caption{{Same as figure~\ref{Ising-theory-plots-1}. These OPE coefficients are more sensitive to the choice of derivatives in the cost function and therefore we are not able to accurately determine them.}}
\label{Ising-theory-plots-2}
\end{figure}

The average values of $\Delta_{\sigma}$ and our determinations of the unknown OPE coefficients $\lambda$ and their standard deviations are given in table~\ref{3dIsingtable} and shown in figure~\ref{Ising-theory-plots-1} and figure~\ref{Ising-theory-plots-2}.
\begin{table}[H]\center
	\begin{tabular}{l|l|}
	\cline{2-2}
	                                                & truncation    \\ \hline
	\multicolumn{1}{|l|}{$\Delta_{\sigma}$}         & 0.518(2)    \\ \hline
	\multicolumn{1}{|l|}{$\lambda^{0}_{T\epsilon T}$} & 0.81(5)     \\
	\multicolumn{1}{|l|}{$\lambda^{0}_{T\epsilon C}$} & 0.30(6)       \\
	\multicolumn{1}{|l|}{$\lambda^{4}_{C\epsilon C}$} & -0.3(1)      \\ \hline
	\multicolumn{1}{|l|}{$\lambda^{3}_{C\epsilon C}$} & -2(2)        \\
	\multicolumn{1}{|l|}{$\lambda^{2}_{C\epsilon C}$} & 2(5)          \\
	\multicolumn{1}{|l|}{$\lambda^{1}_{C\epsilon C}$} & -5(11)          \\
	\multicolumn{1}{|l|}{$\lambda^{0}_{C\epsilon C}$} & -3(11)          \\ 
	\multicolumn{1}{|l|}{$\lambda_{\epsilon' \epsilon \epsilon'}$} & 1(3)      \\ 
	\multicolumn{1}{|l|}{$\lambda_{\epsilon' \epsilon C}$} & 0(2)             \\ \hline
	\end{tabular}
\caption{$\Delta_{\sigma}$ and the unknown OPE coefficients in the 3d Ising CFT found by minimizing $f_{\{r_i\}}(\Delta_{\sigma},\lambda)$ and averaging over the obtained values for different values of the randomly generated numbers $r_i$ and different choices of ${\rm dim\,}\mathcal{D}$.
The OPE coefficients in the bottom part of the table have significantly larger standard deviations than the OPE coefficients in the top part, indicating that we are not able to compute these very accurately using this truncation.  
}\label{3dIsingtable}
\end{table}

All error bars in table~\ref{3dIsingtable} are again estimated using the standard deviation of our averaging procedure, in the same way as in the free theory; hence, it is expected that some of them could be underestimated. Presumably, the reason why some OPE coefficients again have significantly larger error bars than others is the fact that the contributions of these operators are of the same order of magnitude as the truncated part of the correlator. 

We remark that the bound for the OPE coefficient $\lambda^0_{T\epsilon T}$ given in \cite{Cordova:2017zej}  assumes the following form in our normalization:\footnote{The OPE coefficients of the stress tensor in this paper are for the unit-normalized stress tensor.}
\begin{equation}
|\lambda^0_{T\epsilon T}|\leqslant 0.9810,
\end{equation}
It is reassuring to observe that the value we find indeed satisfies this bound.

We can now see what happens if we treat the spin-4 operator of dimension $\Delta_C = 5.022665$ approximately as a conserved current, which may be a reasonable approximation due to its small anomalous dimension. In this approximation, we relate the OPE coefficients $\lambda_{C\epsilon C}^{n_{IJ}}$ for $n_{IJ}=0,1,2,3$ to $\lambda_{C\epsilon C}^{4}$ using the Ward identities. These Ward identities in $d=3$ are given by:
\begin{equation}\label{Ward-Ising}
\begin{split}
&\lambda_{C\epsilon C}^{3}= \frac{4 \Delta _{\epsilon } \left(\Delta _{\epsilon } \left(\left(\Delta _{\epsilon }-16\right) \Delta _{\epsilon }+72\right)-78\right)}{\left(\Delta _{\epsilon }-10\right) \Delta _{\epsilon } \left(\left(\Delta _{\epsilon }-10\right) \Delta _{\epsilon }+32\right)+210}\lambda_{C\epsilon C}^{4} ,\\
& \lambda_{C\epsilon C}^{2}= \frac{\Delta _{\epsilon } \left(\Delta _{\epsilon }+2\right) \left(\Delta _{\epsilon } \left(5 \Delta _{\epsilon }-62\right)+162\right)}{\left(\Delta _{\epsilon }-10\right) \Delta _{\epsilon } \left(\left(\Delta _{\epsilon }-10\right) \Delta _{\epsilon }+32\right)+210} \lambda_{C\epsilon C}^{4},\\
& \lambda_{C\epsilon C}^{1}= \frac{2 \left(\Delta _{\epsilon }-8\right) \Delta _{\epsilon } \left(\Delta _{\epsilon }+2\right) \left(\Delta _{\epsilon }+4\right)}{\left(\Delta _{\epsilon }-10\right) \Delta _{\epsilon } \left(\left(\Delta _{\epsilon }-10\right) \Delta _{\epsilon }+32\right)+210} \lambda_{C\epsilon C}^{4},\\
& \lambda_{C\epsilon C}^{0}= \frac{\Delta _{\epsilon } \left(\Delta _{\epsilon }+2\right) \left(\Delta _{\epsilon }+4\right) \left(\Delta _{\epsilon }+6\right)}{8 \left(\left(\Delta _{\epsilon }-10\right) \Delta _{\epsilon } \left(\left(\Delta _{\epsilon }-10\right) \Delta _{\epsilon }+32\right)+210\right)} \lambda_{C\epsilon C}^{4},\\
&\lambda_{\epsilon' \epsilon C}=0.
\end{split}
\end{equation}

Again, the sets of derivatives are chosen such that $\Delta_{\sigma}$ is closest to the exact value with the minimal standard deviation. Here we scan over the set of the equations~(\ref{equations}) obtained by taking three or fewer derivatives with respect to the cross-ratios $(a^{\pm},b^{\pm},w)$. The sets of derivatives that we use are given by eq.~(\ref{isingwardder}).  We present the results in table~\ref{3dIsingtableWard}.

\begin{table}[h!]\center
\begin{tabular}{l|l|llll}
\cline{2-2} \cline{6-6}
                                                               & truncation &    &                     & \multicolumn{1}{l|}{}                          & \multicolumn{1}{l|}{eq.~(\ref{Ward-Ising})} \\ \cline{1-2} \cline{5-6} 
\multicolumn{1}{|l|}{$\Delta_\sigma$}                          &     0.518(1)       &   & \multicolumn{1}{l|}{} & \multicolumn{1}{l|}{$\lambda_{C\epsilon C}^3$} & \multicolumn{1}{l|}{-0.43(2)}         \\ \cline{1-2}
\multicolumn{1}{|l|}{$\lambda_{T\epsilon T}^0$}                &    0.83(8)        &   & \multicolumn{1}{l|}{} & \multicolumn{1}{l|}{$\lambda_{C\epsilon C}^2$} & \multicolumn{1}{l|}{5.7(2)}         \\
\multicolumn{1}{|l|}{$\lambda_{T\epsilon C}^0$}                &     0.44(7)        &   & \multicolumn{1}{l|}{} & \multicolumn{1}{l|}{$\lambda_{C\epsilon C}^1$} & \multicolumn{1}{l|}{-4.8(2)}         \\
\multicolumn{1}{|l|}{$\lambda_{C\epsilon C}^4$}                &     -0.44(2)       &  & \multicolumn{1}{l|}{} & \multicolumn{1}{l|}{$\lambda_{C\epsilon C}^0$} & \multicolumn{1}{l|}{0.34(2)}         \\ \cline{1-2} \cline{5-6} 
\multicolumn{1}{|l|}{$\lambda_{\epsilon' \epsilon \epsilon'}$} &    3(4)        &  &                       &                                                &                                  \\ \cline{1-2}
\end{tabular}
\caption{The left table shows $\Delta_{\sigma}$ and the unknown OPE coefficients in the 3d critical Ising CFT found by minimizing $f_{\{r_i\}}(\Delta_{\sigma},\lambda)$ and averaging over the values obtained for different sets of randomly generated numbers $r_i$ and different choices of ${\rm dim\,}\mathcal{D}$. These computations impose the Ward identity constraint (\ref{Ward-Ising}), which is not exact but may be approximately satisfied in the theory. The $\lambda_{\epsilon' \epsilon \epsilon'}$ OPE coefficient has a large standard deviation and we are not able to compute this coefficient accurately. The table on the right shows the values of the OPE coefficients $\lambda_{C\epsilon C}^{n_{IJ}}$ for $n_{IJ}=0,1,2,3$ computed using the Ward identity (\ref{Ward-Ising}) and the value of $\lambda_{C\epsilon C}^4$ given in the left table.
}\label{3dIsingtableWard}
\end{table}

Recently, the OPE coefficient $\lambda_{T\epsilon T}$ was computed in \cite{Hu:2023xak}, using the fuzzy sphere regularization developed in \cite{Zhu:2022gjc, Hu:2023xak}. The value they obtained is $\lambda_{T\epsilon T} \approx 0.8658(69)$. To directly compare their results with ours, one must first ensure that they are written in the same basis of 3-point tensor structures. The relation between this OPE coefficient and the $\lambda_{T\epsilon T}^{n_{IJ}}$ coefficients in the standard box basis is given by\footnote{We thank Zheng Zhou and Yin-Chen He for pointing out an error in the previous version of this formula.}
\begin{equation}
\lambda_{T\epsilon T}=\frac{2}{15} \lambda_{T\epsilon T}^{0} - \frac13 \lambda_{T\epsilon T}^{1} + \lambda_{T\epsilon T}^{2}\,.
\end{equation}
Using the Ward identities for the $\lambda_{T\epsilon T}^{n_{IJ}}$ coefficients (see e.g.~eq.~(4.123) in \citep{Poland:2021xjs}) one finds
\begin{equation}
\lambda_{T \epsilon T} = \frac{4(\Delta_{\epsilon} - 3)(\Delta_{\epsilon} - 5)}{5 \Delta_{\epsilon} (\Delta_{\epsilon} + 2)} \lambda_{T\epsilon T}^{0}\approx 0.77(5),
\end{equation}
where we use $\lambda_{T\epsilon T}^{0} \approx 0.81(5)$. This confirms that our result is consistent with the one in \cite{Hu:2023xak} to within two standard deviations.

\newpage
\section{Discussion}
\label{sec:discussion}

In this paper we studied five-point correlation functions of scalar operators in conformal field theories. We developed a new method for computing the five-point conformal blocks by using an appropriate ansatz in radial coordinates and solving two quadratic Casimir differential equations perturbatively. The ansatz that we write down works for arbitrary spin of the exchanged operators and for arbitrary intermediate conformally-invariant tensor structures. It is also applicable in an arbitrary number of spatial dimensions $d>2$. We believe that this ansatz can be straightforwardly generalized to six-point conformal blocks in the snowflake channel. An open question for future research is whether an analogous ansatz can be constructed for conformal blocks in the comb channel of six- and higher-point correlators.

In principle, we could directly derive recursion relations for the coefficients $c(m_1,m_2,j_1,j_2,k)$, similar to the recursion relations derived for $c$-coefficients in the radial expansion of the four-point conformal blocks \cite{Costa:2016xah}. The recursion relations could be obtained by demanding that the ansatz (\ref{blocks-radial}) satisfy the quadratic Casimir equations at arbitrary order in $r_1$ and $r_2$. However, such recursion relations would contain a couple thousand terms, so that solving them would not be significantly easier than solving the Casimir differential equations perturbatively.

In this work we also computed the OPE coefficients of all operators (including subleading twists) that contribute to $\langle \phi \phi [\phi,\phi]_{0,0}\phi\phi\rangle $ in mean-field theories. A possible straightforward generalization would be to consider the mean-field theory correlator $\langle\phi_1 \phi_2 [\phi_2,\phi_2]_{0,0} \phi_2 \phi_1 \rangle$ or some of its variations. One can also use the lightcone bootstrap (or perhaps an inversion formula) to compute corrections to the subleading OPE coefficients at large spin in interacting theories, generalizing the approach of~\cite{Antunes:2021kmm}.

Using our result, we first performed a numerical study of five-point correlators in the free scalar theory (as a testing ground) and then of the critical 3d Ising model. In particular, we truncated the OPEs in the five-point correlators and demanded  that the truncated correlators approximately satisfy the crossing relation in order to compute the OPE coefficients between one scalar and two spinning operators. An advantage of this approach to the conformal bootstrap is that it does not require positivity. Hence, it is possible to apply it to five-point correlators, which do not exhibit a positivity property, unlike four-point functions. The main disadvantage is the fact that it is difficult to accurately estimate the error bars of the results, as this in general requires us to accurately parameterize the truncated part of the correlator, which is difficult. However, our overall conclusion is that the five-point bootstrap works and can be used to determine OPE coefficients which may be more difficult to study using spinning four-point correlators. 

The success of our study provides motivation to add more contributions to the truncated correlators, or in other words, to increase the value of $\Delta_{\rm cutoff}$. In the free theory we attempted to add some of the contributions of the spin-6 operator, specifically, $([\phi,\phi]_{0,0}, [\phi,\phi]_{0,6})$ and $([\phi,\phi]_{0,2}, [\phi,\phi]_{0,6})$. We observed that these contributions do not push $e_{i}(\Delta_{\phi}=1/2,\lambda)$ closer to zero when it is evaluated using the known free theory OPE coefficients $\lambda$, given by eq.~(\ref{antunes}). Therefore, we concluded that it is not consistent to include just these spin-6 contributions without also adding $([\phi,\phi]_{0,4}, [\phi,\phi]_{0,6})$ and $([\phi,\phi]_{0,6}, [\phi,\phi]_{0,6})$. However, in our initial attempts we found that computing all of the spin-6 contributions was fairly time-consuming, providing some motivation for improving the block computation algorithm. Namely, one needs a more efficient algorithm for computing the five-point conformal blocks and their derivatives for the exchanged operators of spin six and higher. Another idea potentially worth exploring in the future is to approximate the truncated operators in the OPEs using their mean-field theory counterparts with MFT conformal dimensions and OPE coefficients. One could also explore the use of stochastic approaches to minimizing the cost function~\cite{Kantor:2021kbx, Kantor:2021jpz, Laio:2022ayq, Kantor:2022epi}, which may become particularly useful as $\Delta_{\rm cutoff}$ is increased.

It could also be fruitful to extend our study to other OPEs in the critical 3d Ising model. For example, by examining the five-point correlator $\langle \sigma \epsilon \epsilon \epsilon \sigma\rangle$ in the same OPE channels as considered in this paper, one would find OPE coefficients of one scalar operator $(\epsilon)$ and two spinning, $\mathbf{Z}_2$-odd operators.  Moreover, one could obtain additional constraints from the $\langle \sigma \sigma \epsilon \sigma \sigma \rangle$ correlator by considering the $\sigma \times \epsilon$ OPE in one of the channels. This relation to the $\mathbf{Z}_2$-odd spectrum might improve some of the OPE coefficient predictions given in this paper.

\subsection*{Acknowledgments}
We thank Alexandre Belin, Aleksandar Bukva, Rajeev Erramilli, Walter Goldberger, Yin-Chen He, Murat Koloğlu, Matthew Mitchell, Vasilis Niarchos, Costis Papageorgakis, Slava Rychkov, Witold Skiba, Ning Su, Yuan Xin, and Zheng Zhou for discussions. The work of D.P. and P.T. is supported by U.S. DOE grant DE-SC00-17660 and Simons Foundation grant 488651 (Simons Collaboration on the Nonperturbative Bootstrap). The work of V.P. is supported by the Perimeter Institute for Theoretical Physics. Computations in this work were performed on the Yale Grace computing cluster, supported by the facilities and staff of the Yale University Faculty of Sciences High Performance Computing Center.

\appendix

\section{Exact OPE coefficients in the free theory}
\label{app:exact}

Here, we  give the exact expressions for the OPE coefficients $\lambda^{0}_{T\phi^2 T}$, $\lambda^{0}_{T\phi^2 C}$, and $\lambda^{n_{IJ}}_{C\phi^2 C}$ in the free theory, as calculated in \cite{Antunes:2021kmm}:
\begin{equation}
\begin{split}
& \lambda^{0}_{T\phi^2 T}=\frac{3}{4 \sqrt{2}},\quad \lambda^{0}_{T\phi^2 C}= \frac{1}{32}\sqrt{\frac{105}{2}},\quad \lambda^{4}_{C\phi^2 C}=\frac{1}{32 \sqrt{2}},\\
& \lambda^{3}_{C\phi^2 C}=-\frac{7}{8 \sqrt{2}},\quad \lambda^{2}_{C\phi^2 C}= \frac{105}{32 \sqrt{2}},\quad \lambda^{1}_{C\phi^2 C}= -\frac{35}{16 \sqrt{2}},\quad \lambda^{0}_{C\phi^2 C}=\frac{35}{256 \sqrt{2}}.
\end{split}
\end{equation}

\section{Sets of derivatives in the free theory}\label{free-theory-derivatives}

Here we give the sets of derivatives that we use when computing the OPE coefficients in the free theory:
\begin{equation}\label{ftd}
\begin{split}
\mathcal{D}&=\{\partial_{w}, \partial_{b^{+}},  \partial_{w}\partial_{b^{+}},  \partial_{w}^2\partial_{b^{+}},\partial_{b^{+}}^2, \partial_{w}\partial_{b^{+}}^2, \partial_{a^{+}}^2, \partial_{b^{-}}^2  \},\\
\mathcal{D}&=\{\partial_{w}, \partial_{w}^2, \partial_{w}^3, \partial_{b^{+}}, \partial_{w}\partial_{b^{+}}, \partial_{a^{+}}^2, \partial_{b^{+}}^2, \partial_{b^{+}}^3, \partial_{b^{-}}^2\partial_{b^{+}} \},\\
\mathcal{D}&=\{\partial_{w}, \partial_{w}^2, \partial_{w}^3, \partial_{b^{+}}, \partial_{w}\partial_{b^{+}}, \partial_{w}^2\partial_{b^{+}}, \partial_{b^{+}}^2, \partial_{b^{+}}^3, \partial_{a^{+}}^2, \partial_{b^{-}}^2\partial_{w}\},\\
\mathcal{D}&=\{\partial_{w}, \partial_{w}^2, \partial_{w}^3, \partial_{b^{+}}, \partial_{w}\partial_{b^{+}}, \partial_{w}^2\partial_{b^{+}}, \partial_{b^{+}}^2, \partial_{b^{+}}^3, \partial_{a^{+}}^2, \partial_{b^{-}}^2\partial_{w}, \partial_{b^{-}}^2\partial_{b^+}\},\\
\mathcal{D}&=\{\partial_{w}, \partial_{w}^2, \partial_{w}^3, \partial_{b^{+}}, \partial_{w}\partial_{b^{+}}, \partial_{w}^2\partial_{b^{+}}, \partial_{b^{+}}^2, \partial_{b^{+}}^3, \partial_{a^{+}}^2, \partial_{a^{+}}^2\partial_{w}, \partial_{b^{-}}^2\partial_{w}, \partial_{b^{-}}^2\partial_{b^+}\},\\
\mathcal{D}&=\{\partial_{w}, \partial_{w}^2, \partial_{w}^3, \partial_{b^{+}}, \partial_{w}\partial_{b^{+}}, \partial_{w}^2\partial_{b^{+}}, \partial_{b^{+}}^2, \partial_{b^{+}}^2\partial_{w}, \partial_{b^{+}}^3, \partial_{a^{+}}^2, \partial_{a^{+}}^2\partial_{w}, \partial_{b^{-}}^2\partial_{w}, \partial_{b^{-}}^2\partial_{b^+}\},\\
\mathcal{D}&=\{\partial_{w}, \partial_{w}^2, \partial_{b^{+}}, \partial_{w}\partial_{b^{+}}, \partial_{w}^2\partial_{b^{+}}, \partial_{b^{+}}^2, \partial_{b^{+}}^2\partial_{w}, \partial_{b^{+}}^3, \partial_{a^{+}}^2, \partial_{a^{+}}^2\partial_{w}, \partial_{b^{-}}^2\partial_{w}, \partial_{b^{-}}^2, \partial_{b^{-}}^2\partial_{b^+}, \partial_{a^{-}}^2\partial_{w}\}.
\end{split}
\end{equation}

The sets of derivatives that we use when we impose the Ward identity for the spin-4 OPE coefficients are given by:
\begin{equation}\label{ftdtwo}
\begin{split}
\mathcal{D}&=\{\partial_{w}, \partial_{w}\partial_{b^{-}}^2, \partial_{w}\partial_{a^{-}}^2, \partial_{a^+}^2\},\\
\mathcal{D}&=\{\partial_{w}, \partial_{w}\partial_{b^+},  \partial_{w}^2\partial_{b^+}, \partial_{w}\partial_{b^+}^2, \partial_{w}\partial_{a^{-}}^2\},\\
\mathcal{D}&=\{\partial_{w}, \partial_{w}\partial_{b^+},  \partial_{w}^2\partial_{b^+}, \partial_{w}\partial_{b^+}^2, \partial_{w}\partial_{a^{-}}^2, \partial_{w}\partial_{b^{-}}^2\}.
\end{split}
\end{equation}

\section{Sets of derivatives in the critical 3d Ising model}\label{Ising-derivatives}

Here, we give the sets of derivatives that we use when computing the OPE coefficients in the critical Ising model:
\begin{equation}\label{imd}
\begin{split}
\mathcal{D}=\{&\partial_{w}, \partial_{w}^2, \partial_{w}^3, \partial_{b^+}, \partial_{b^+}\partial_{w}, \partial_{b^+}\partial_{w}^2, \partial_{a^+}^2, \partial_{a^+}^2\partial_{w}, \partial_{b^+}^2, \partial_{b^+}\partial_{b^-}^2   \},\\
\mathcal{D}=\{&\partial_{w}, \partial_{w}^2, \partial_{w}^3, \partial_{b^+}, \partial_{b^+}\partial_{w}, \partial_{b^+}\partial_{w}^2, \partial_{a^+}^2\partial_{w},  \partial_{b^+}^2, \partial_{b^+}^3,  \partial_{b^+}\partial_{b^-}^2, \partial_{b^-}^2\},\\
\mathcal{D}=\{&\partial_{w}, \partial_{w}^2, \partial_{w}^3, \partial_{b^+}, \partial_{b^+}\partial_{w}, \partial_{b^+}\partial_{w}^2, \partial_{a^+}^2, \partial_{a^+}^2\partial_{w}, \partial_{b^{+}}^2\partial_{w}, \partial_{a^{+}}^2\partial_{b^{+}}, \partial_{b^+}^3, \partial_{b^-}^2\},\\
\mathcal{D}=\{&\partial_{w}, \partial_{w}^2, \partial_{w}^3, \partial_{b^+}, \partial_{b^+}\partial_{w}, \partial_{b^+}\partial_{w}^2, \partial_{a^+}^2, \partial_{a^+}^2\partial_{w}, \partial_{b^+}^2, \partial_{b^{+}}^2\partial_{w}, \partial_{a^{+}}^2\partial_{b^{+}}, \partial_{b^+}^3, \partial_{b^-}^2\},\\
\mathcal{D}=\{&\partial_{w}, \partial_{w}^2, \partial_{w}^3, \partial_{b^+}, \partial_{b^+}\partial_{w}, \partial_{a^+}^2, \partial_{a^+}^2\partial_{w}, \partial_{b^+}^2, \partial_{b^{+}}^2\partial_{w}, \partial_{a^{+}}^2\partial_{b^{+}}, \partial_{b^+}^3, \partial_{b^-}^2, \partial_{b^-}^2\partial_{w}, \partial_{a^-}^2\partial_{w} \},\\
\mathcal{D}=\{&\partial_{w}, \partial_{w}^2, \partial_{w}^3, \partial_{b^+}, \partial_{b^+}\partial_{w}, \partial_{a^+}^2, \partial_{a^+}^2\partial_{w}, \partial_{b^+}^2, \partial_{b^{+}}^2\partial_{w}, \partial_{a^{+}}^2\partial_{b^{+}}, \partial_{b^+}^3, \partial_{b^-}^2\partial_{b^+}, \partial_{b^-}^2, \partial_{b^-}^2\partial_{w}, \partial_{a^-}^2\partial_{w} \},\\
\mathcal{D}=\{&\partial_{w}, \partial_{w}^2, \partial_{w}^3, \partial_{b^+}, \partial_{b^+}\partial_{w}, \partial_{b^+}\partial_{w}^2, \partial_{a^+}^2, \partial_{a^+}^2\partial_{w}, \partial_{b^+}^2, \partial_{b^{+}}^2\partial_{w}, \partial_{a^{+}}^2\partial_{b^{+}}, \partial_{b^+}^3,\\
& \partial_{b^-}^2\partial_{b^+}, \partial_{b^-}^2, \partial_{b^-}^2\partial_{w}, \partial_{a^-}^2\partial_{w} \}.
\end{split}
\end{equation}

The sets of derivatives that we use when we impose the Ward identity for the spin-4 OPE coefficients are given by:
\begin{equation}\label{isingwardder}
\begin{split}
\mathcal{D}&=\{ \partial_{w}, \partial_{w}\partial_{a^{-}}^2, \partial_{b^+}, \partial_{b^{+}}^2, \partial_{b^{-}}^2 \},\\
\mathcal{D}&=\{ \partial_{w}, \partial_{w}^2, \partial_{a^+}^2, \partial_{a^+}^2\partial_{w}, \partial_{b^-}^2\partial_{w}, \partial_{a^+}^2\partial_{b^+}\},\\
\mathcal{D}&=\{ \partial_{w}, \partial_{w}^2, \partial_{w}^3, \partial_{a^+}^2, \partial_{a^+}^2\partial_{w}, \partial_{b^-}^2\partial_{w}, \partial_{a^+}^2\partial_{b^+}\}.\\
\end{split}
\end{equation}

\bibliographystyle{JHEP}
\bibliography{refs-fivept.bib}{}
\end{document}